\documentstyle[12pt]{article}
\newcounter{subequation}[equation]

\makeatletter

\def\thesubequation{\theequation\@alph\c@subequation}
\def\@subeqnnum{{\rm (\thesubequation)}}
\def\slabel#1{\@bsphack\if@filesw {\let\thepage\relax
   \xdef\@gtempa{\write\@auxout{\string
      \newlabel{#1}{{\thesubequation}{\thepage}}}}}\@gtempa
   \if@nobreak \ifvmode\nobreak\fi\fi\fi\@esphack}
\def\subeqnarray{\stepcounter{equation}
\let\@currentlabel=\theequation\global\c@subequation\@ne
\global\@eqnswtrue
\global\@eqcnt\z@\tabskip\@centering\let\\=\@subeqncr
$$\halign to \displaywidth\bgroup\@eqnsel\hskip\@centering
  $\displaystyle\tabskip\z@{##}$&\global\@eqcnt\@ne
  \hskip 2\arraycolsep \hfil${##}$\hfil
  &\global\@eqcnt\tw@ \hskip 2\arraycolsep
  $\displaystyle\tabskip\z@{##}$\hfil
   \tabskip\@centering&\llap{##}\tabskip\z@\cr}
\def\endsubeqnarray{\@@subeqncr\egroup
                     $$\global\@ignoretrue}
\def\@subeqncr{{\ifnum0=`}\fi\@ifstar{\global\@eqpen\@M
    \@ysubeqncr}{\global\@eqpen\interdisplaylinepenalty \@ysubeqncr}}
\def\@ysubeqncr{\@ifnextchar [{\@xsubeqncr}{\@xsubeqncr[\z@]}}
\def\@xsubeqncr[#1]{\ifnum0=`{\fi}\@@subeqncr
   \noalign{\penalty\@eqpen\vskip\jot\vskip #1\relax}}
\def\@@subeqncr{\let\@tempa\relax
    \ifcase\@eqcnt \def\@tempa{& & &}\or \def\@tempa{& &}
      \else \def\@tempa{&}\fi
     \@tempa \if@eqnsw\@subeqnnum\refstepcounter{subequation}\fi
     \global\@eqnswtrue\global\@eqcnt\z@\cr}
\let\@ssubeqncr=\@subeqncr
\@namedef{subeqnarray*}{\def\@subeqncr{\nonumber\@ssubeqncr}\subeqnarray}

\@namedef{endsubeqnarray*}{\global\advance\c@equation\m@ne%
                           \nonumber\endsubeqnarray}

\makeatletter
\@addtoreset{equation}{section}
\makeatother
\renewcommand{\theequation}{\thesection.\arabic{equation}}

\date{}
\begin{document}

\begin{titlepage}

\begin{center}
\hfill hep-th/0007086\\
\hfill UM--TH/00-10\\
\hfill OHSTPY-HEP-T-00-011\\

\vskip 2.5 cm
{\Large \bf Conformal Sigma Models for a Class of $T^{p,q}$ Spaces}

\vskip 1 cm

{\large L.A. Pando Zayas$^a$ and A.A. Tseytlin$^{b,}$\footnote{Also at
 Imperial College, London and Lebedev
Institute, Moscow.}}\\

\end{center}

\vskip 1cm
\centerline{\it ${}^a$ Randall Laboratory of Physics,
The University of Michigan}
\centerline{\it Ann Arbor, Michigan 48109-1120, USA}
\centerline{\tt lpandoz@umich.edu}

\vskip 0.8 cm

\centerline{\it ${}^b$ Department of Physics, The Ohio State
University,}
\centerline{\it 174 West 18th Avenue, Columbus, OH 43210, USA}
\centerline{\tt tseytlin@mps.ohio-state.edu }

\vskip 1.5 cm

\begin{abstract}
We consider  a 2-d conformal theory
  based on  $G \times G' \over H$ coset sigma model
 introduced  by Guadagnini, Martellini and Mintchev.
 It is  shown that in the case of
 $SU(2) \times SU(2) \over U(1)$
 the metric of the corresponding
 background is of $T^{p,q}$  coset space form (but  is not
  an Einstein one).
 Similar interpretation is possible for the
 Lorentzian  coset space $ W_{4,2}= {SL(2,R) \times SL(2,R) \over U(1)}$.
 The resulting 10-d homogeneous space  metric on  $W_{4,2}\times T^{p,q}$
 supplemented with   2-form   field
  gives  a critical  NS-NS  superstring background
 with  conformal sigma model interpretation.
\end{abstract}

\end{titlepage}

\def\o{\omega}
\def\O{\Omega}
\def\e{\epsilon}
\def\pd{\partial}
\def\pdz{\partial_{\bar{z}}}
\def\bz{\bar{z}}
\def\e{\epsilon}
\def\m{\mu}
\def\n{\nu}
\def\a{\alpha}
\def\b{\beta}
\def\g{\gamma}
\def\G{\Gamma}
\def\d{\delta}
\def\r{\rho}
\def\bx{\bar{x}}
\def\by{\bar{y}}
\def\bm{\bar{m}}
\def\bn{\bar{n}}
\def\s{\sigma}
\def\na{\nabla}
\def\D{\Delta}
\def\l{\lambda}
\def\te{\theta}
\def\na{\bigtriangledown}
\def\p{\phi}
\def\L{\Lambda}
\def\hR{\hat R}
\def\ch{{\cal H}}
\def\ep{\epsilon}
\def\bj{\bar{J}}

\section{Introduction}

Metrics of physically interesting backgrounds  usually  have large
amount of  global symmetry.
 One set  of  examples are black  hole metrics with
   rotational symmetry,
 and another  are symmetric   spaces
    $AdS_n\times S^{n}$  supported  by R-R antisymmetric tensor
    backgrounds.
At the same time,   string  solutions
which have   known  2-d CFT  interpretation,
like gauged WZW models,   have
associated  space-time metrics with very few  or no global symmetries.
 It is of interest to look for new examples
 of conformal sigma models  related to
 metrics  on  symmetric spaces.

Special   symmetric spaces that were   recently
 discussed in connection with  AdS/CFT
correspondence  are $T^{p,q}$ spaces. These are cosets
of the form $T^{p,q}=[SU(2)\times SU(2)]/U(1)$ with the  integers
$p$ and $q$ determining
the embedding of the $U(1)$  subgroup. Their metric is
      \cite{pope,romans,candelas}
\begin{equation}
ds^2=\l_1^2(d\te_1^2+\sin^2 \te_1d\p_1^2)
+\l_2^2(d\te_2^2+\sin^2 \te_2d\p_2^2) \nonumber \\
+\l^2(d\psi+p\cos \te_1 d\p_1 +q\cos\te_2 d\p_2)^2.
\end{equation}
A particular case of   $p=q=1$  relevant for discussions of  AdS
supergravity  solutions  preserves part
of  supersymmetry, and  with  $\l_1^2=\l_2^2=1/6$
and $\l^2=1/9$  its metric is an   Einstein space  one.

Below we shall show that  certain   symmetric metrics  on
$T^{p,q}$ spaces  can be interpreted as parts of
NS-NS string backgrounds  associated  with  a class of   conformal
coset sigma models  proposed by Guadagnini,
Martellini and Mintchev (GMM)  \cite{gmm,gmmcft}.
Though  these metrics  supported by NS-NS 2-form field
 are not Einstein ones  (in contrast to the $T^{1,1}$ example
  studied  \cite{tpqbrane} in
connection with AdS/CFT correspondence),
they may still turn out to be of some interest.

The conformal sigma model  with   $T^{p,q}$ metric
should be supplemented by another  one to balance the central charge.
One example of a critical $D=10$ string model
has  the  metric of the
form $W_{4,2}\times T^{1,q}$, where
$W_{4,2}=SO(2,2)/SO(2)=[SL(2,R)\times SL(2,R)]/U(1)$.
An Einstein
representative in the class of metrics on  $W_{4,2}$
 was discussed in \cite{stro} as a
generalization of $AdS_5$.

In section 2 we shall review the GMM
construction \cite{gmm}
and its interpretation as a coset CFT
\cite{gmmcft}. We shall also  comment on its relation
to a class of  gauged WZW models as discussed in  \cite{wholo}.
In section 3
we shall  consider  a GMM model
that leads to a   $T^{p,q}$ metric.
 Section 4 is devoted
to  explicit check of conformal invariance
of the corresponding sigma model at the one- and two-loop
levels.

 

\section{The Guadagnini-Martellini-Mintchev Model}

The starting point is the WZW action    \cite{wzw}
\begin{equation}
I(U;n)=\frac{n}{8\pi}\int\limits_{\partial  M} d^2x\ Tr (\pd_\mu
U\pd^\mu U^{-1} )
+\frac{n}{12\pi}\int\limits_{M}d^3y\ \epsilon^{ijk}\  Tr (
U^{-1}\pd_iUU^{-1}\pd_jUU^{-1}\pd_kU) ,
\end{equation}
where  $U$ is an element of the group $G$ and  $n$ is the level of the
associated affine Kac-Moody algebra. The
property of the WZW model that is used in the  GMM   construction
is that under an arbitrary variation of the group element $\d U$
the WZW action changes by
\begin{equation}
\label{left}
\d I(U;n)=\frac{n}{4\pi}\int d^2x\ Tr[  U^{-1}\d
U(\eta_{\m\n}-\e_{\m\n})
\pd_\m(U^{-1}\pd_\n U)].
\end{equation}
This variation can be  written also as
\begin{equation}
\label{right}
\d I(U;n)=\frac{n}{4\pi}\int d^2x\ Tr[\d U U^{-1}(\eta_{\m\n}+\e_{\m\n})
\pd_\m(\pd_\n UU^{-1})],
\end{equation}
or as  $ \int d^2z\ Tr[U^{-1}\d
U\pd_{z}(U^{-1}\pdz U)]= \int d^2z  \
Tr[ \d UU^{-1}\pdz(\pd_z UU^{-1})]$.
 From these variations one can read
off the
currents associated with the symmetry
$U\to \O(z)U\bar{\O}^{-1}(\bar{z})$ \cite{wzw}.

Consider the variation of the WZW model under the
following gauge transformation
\begin{equation}
\label{tra1}
U\to U R(\Omega^{-1}),
\end{equation}
where $R$ is a representation of a subgroup $H\subset G$ and
$\Omega \in H$. Under
infinitesimal transformations $\Omega(x)=1+\omega(x)$ the WZW action
transforms as (\ref{left})
\begin{equation}
\delta I(U;n)=-\frac{n}{4\pi}\int d^2 \ Tr  [
R(\o)\pd^\m(U^{-1}\pd_\m U-\e^{\m\n}U^{-1}\pd_\n U)],
\end{equation}
where we set  $R(\Omega^{-1}) = 1 - R(\omega)+ ...$.
In order to cancel this
``classical anomaly"  GMM  suggested to introduce  another
 field $V$ belonging to a group $ G'$  whose action has  similar
anomalous
transformation property under $H$.
It is assumed that the same $H$ is a subgroup of both $G$ and $G'$
so that the class of resulting coset models is rather special.
Let  $V\in G'$
and   $R'$ be a representation of $H\subset G'$ acting on $V$ according
to
\begin{equation}
\label{tra2}
V\to R'(\O)V.
\end{equation}
Using Eq. (\ref{right}) we get
for the variation of the WZW action $         I(V;m)$
similar to (2.1)
\begin{equation}
\delta I(V;m)=\frac{m}{4\pi}\int d^2 \ Tr[ R'(\o)\pd^\m(\pd_\m VV^{-1}
+\e_{\m\n}\pd^\n VV^{-1})].
\end{equation}
One can then check that the model
\begin{eqnarray}
\label{gmmm}
I_{GMM}&=&I(U;n)+I(V;m)+I_{int}(U,V;k)\ ,   \nonumber \\
I_{int}(U,V;k)&=&-\frac{k}{2\pi}\int d^2x \left[Tr (R_\a U^{-1}\pd_\m
U)Tr
(R'_\a\pd^\m VV^{-1})\right. \nonumber \\
&+&\e^{\m\n}Tr (R_\a U^{-1}\pd_\m U)\left. Tr
(R'_\a\pd_\n VV^{-1})\right]
\end{eqnarray}
is gauge invariant for
\begin{eqnarray}
\label{condmn}
n&=&kr',\qquad  m=kr, \nonumber \\
Tr R_\a R_\b&=&r\delta_{\a\b}, \qquad Tr R'_\a R'_\b=r'\delta_{\a\b},
\end{eqnarray}
where, as in
\cite{gmm}, the generators of the Lie algebras  of
${G}$ and
${ G'}$ are $\{R_i\}=\{R_I, R_\a\}$ and $\{R'_a\}=\{R'_A, R'_\a\}$,
where
$R_\a$ and $R'_\a$ correspond to the Lie algebra of subgroup  $H$.
The one-loop finiteness of this model was checked in \cite{gmm} and
finiteness at the two-loop level was checked in \cite{belo}.
The conformal field theory defined by
 this sigma model was  discussed  in   ref.
\cite{gmmcft},  which    found
  the current algebra and the Virasoro algebra
with a central charge value coinciding with that of the GKO
construction  \cite{gko,halpern} for
the coset  $(G\times
G')/H$.

 Let  us  briefly  review the
conformal structure of the GMM model.  The variation of
the action (\ref{gmmm}) with respect to $U$ and $V$ yields the following
equations of motion
\begin{equation}
\pdz J_z^i=0, \qquad \pd_z J^a_{\bz}=0,
\end{equation}
where
\begin{eqnarray}
\label{currents}
J^i_z&=&(\pd_z UU^{-1})^i+\frac{1}{r'}(UR_\a
U^{-1})^i
Tr(R'_\a\pd_z VV^{-1}), \nonumber \\
J^a_{\bz}&=&-(V^{-1}\pdz V)^a+\frac{1}{r}(V^{-1}
R'_\a V)^a
Tr(R_\a U^{-1}\pdz U).
\end{eqnarray}
The form  of the equations of motion and  currents suggests, by
analogy with the WZW model, the existence of two copies of affine
 algebras \cite{gmmcft}. Introducing
\begin{equation}
K_z^a=(\pd_zVV^{-1})^{a}, \qquad
K_{\bz}^i=-(U^{-1}\pd_{\bz}U)^i,
\end{equation}
one can write the components of the classical energy-momentum tensor as
\begin{eqnarray}
T_z&=&{1\over kr'}J_z^iJ_z^i+\frac{1}{kr}K_z^AK_z^A, \nonumber \\
T_{\bz}&=&\frac{1}{kr}J_{\bz}^aJ_{\bz}^a+\frac{1}{kr'}K_{\bz}^IK_{\bz}^I.
\end{eqnarray}
The analysis of this  bosonic
 model at the quantum level reveals that the central
charge is  \cite{gmmcft}
\begin{equation}
\label{cc}
c_{GMM}=c(G,kr')+c(G',kr)-c(H,2krr'),
\end{equation}
with $c(G,n)=n\,{\rm dim}G/[n+ 2c_V(G)]$, where
$c_V(G)\d_{ab}=f_{acd}f_{bcd}$.
 The quantum energy-momentum tensor
is of the same form as the classical one but with rescaled coefficients
 \cite{gmmcft,kz}
\begin{eqnarray}
\label{virasoro}
T_z&=&\frac{1}{kr'+2c_V(G)}:J_z^iJ_z^i:
+\frac{1}{kr+2c_V(G')}:K^A_zK^A_z:,\nonumber \\
T_{\bz}&=&\frac{1}{kr+2c_V(G')}:J_{\bz}^aJ_{\bz}^a:
+\frac{1}{kr'+2c_V(G)}:K_{\bz}^IK_{\bz}^I:.
\end{eqnarray}
The expressions in the supersymmetric case are similar,
with levels shifted  ($kr'+2c_V(G) \to kr'$, etc)
    as in the (gauged) WZW model case (see, e.g., \cite{cftsigma}
    and refs. there).


Let us note also that the  GMM
model can be represented
 as a  kind of generalized
gauged WZW model
which  is free of
anomalies and upon elimination of the 2-d  gauge fields
 reduces to  the GMM action. Introducing non-dynamical 2-d gauge fields
 $A$ and $B$  one may consider the   action
\begin{eqnarray}
\label{ggmmm}
\hat I_{GMM}&=&I(U;n)+I(V;m)+I_{int}(U,V,A,B;k), \nonumber \\
I_{int}(U,V;k)&=&-\frac{k}{4\pi}\int d^2z \left[ Tr (R_\a A_{\bz})
 Tr
(R'_\a\pd_z VV^{-1}) -Tr (R'_\a B_{z}) Tr( R_\a U^{-1}\pd_{\bz} U)
\right.
\nonumber \\
&+&\left. Tr (R_\a A_{\bz})Tr (R'_\a B_{z}) \right],
\end{eqnarray}
which is  invariant under the following gauge transformations:
\begin{eqnarray}
\d U&=&-U\o, \qquad \d V=\o V,  \nonumber \\
\d B_i&=&-\pd_i\o-[B_i,\o], \qquad \d A_i=-\pd_i\o-[A_i,\o].
\end{eqnarray}
 Integrating out  the gauge fields gives
 back  the GMM action Eq.(\ref{gmmm}).\footnote{Note that,
 in contrast to what happens in the usual  gauged WZW models
  \cite{wil,dil},
  integrating out the
gauge fields  gives trivial determinant,
i.e. does not produce a non-constant
dilaton coupling.}
     In the  standard
diagonal vector gauged WZW model the gauge action is
$g \to hgh^{-1}$.
 The GMM model may be interpreted as a gauged WZW model
 defined on
the   product group $G'\times G$, with the gauged
subgroup acting  as  \ $(V,U) \to (hV,Uh^{-1})$,
i.e.  it  may be viewed as   a  non-anomalous ``sum" of
right and left  gauged \cite{wholo} WZW models.

\section{$T^{p,q}$ and  $W_{4,2}$ metrics from GMM model }

Let us consider the GMM model for $G=SU(2)$,\, $G'=SU(2)$,\, and
$H=U(1)$. The
$SU(2)$ group elements are parametrized according to
\begin{eqnarray}
\label{param}
U&=&\exp(i\phi_1\s_3)\exp(i\theta_1\s_2)\exp(i\psi_1\s_3), \nonumber \\
V&=&\exp(i\phi_2\s_3)\exp(i\theta_2\s_2)\exp(i\psi_2\s_3).
\end{eqnarray}
 The gauge action of the
$U(1)$ subgroup is defined by
\begin{equation}
\psi_1\to \psi_1-p\epsilon(z,\bz), \qquad \phi_2\to
\phi_2+q\epsilon(z,\bz).
\end{equation}
This corresponds to gauging the subgroup generated by
$i(q\s_3^L-p\s_3^R)$. Consider the sum of the two
WZW models on $SU(2)$  with levels $k_1$ and $k_2$  and the   GMM interaction
term     (\ref{gmmm})
with coefficient $k_3$
\begin{eqnarray}
I&=&{1\over 4\pi}\int d^2x\left[ k_1\bigg( \pd_\m\theta_1\pd^\m\theta_1
+\pd_\m\phi_1\pd^\m\phi_1+\pd_\m\psi_1\pd^\m\psi_1
+\cos(2\theta_1)\pd_\m\phi_1\pd_\n\psi_1
(\eta^{\m\n}+\epsilon^{\m\n}) \bigg)\right.\nonumber \\
&+&k_2\bigg(\pd_\m\theta_2\pd^\m\theta_2+\pd_\m\phi_2\pd^\m\phi_2
+\pd_\m\psi_2\pd^\m\psi_2
+\cos(2\theta_2)\pd_\m\phi_2\pd_\n\psi_2(\eta^{\m\n}+\epsilon^{\m\n})\bigg)
\nonumber\\
&+&\left.k_3\bigg(\cos(2\theta_1)\pd_\m\phi_1
+\pd_\m\psi_1\bigg)\bigg(\cos(2\theta_2)\pd_\n\psi_2
+\pd_\n\phi_2\bigg)(\eta^{\m\n}+\epsilon^{\m\n}) \right].\label{sens}
\\
\nonumber \end{eqnarray}
For the action to be  invariant under (3.2)
one needs to impose the following algebraic constraints:
\begin{equation}
k_1p=k_3q, \qquad k_2q=k_3p.
\end{equation}
Multiplying these equations we get that
 \begin{equation}
k_3=\sqrt{k_1k_2} \ , \ \ \ \ \ \ \
p/q= \sqrt{k_2/k_1} \ .     \end{equation}
Fixing the gauge as $\p_2=0$ one gets a background whose metric is of
the
(non-Einstein)  $T^{1,Q}$       type
\begin{equation}
ds^2=k[ d\theta_1^2+\sin^2\theta_1d\phi_1^2
+Q^2(d\theta_2^2+\sin^2\theta_2d\phi_2^2)
+(d\psi+\cos\theta_1d\phi_1+Q\cos\theta_2d\phi_2)^2] \ ,
\end{equation}
where we have rescaled all variables by 1/2,
 renamed $\psi_2 \to \p_2, \, \psi_1 \to \psi$
and
introduced
 \begin{equation}
Q=p/q = \sqrt{k_2/k_1}  \ , \ \ \ \ \  \ k=k_1\  .
 \end{equation}
The background also includes the  antisymmetric field
\begin{equation}
B_{\phi_1\psi}=k\cos\theta_1, \quad
B_{\phi_1\phi_2}=kQ\cos\theta_1\cos\theta_2,
\quad B_{\phi_2\psi}=-kQ\cos\theta_2.
\end{equation}
The coefficients in front of the different terms
in the action  are dictated by gauge
invariance of the total action and can not  be re-adjusted.

 Fixing the
gauge in the original variables as $\psi_1=0$ one ends up with a metric
of the type $T^{ Q^{-1},1}$.  More generally,
 imposing  $\psi_1=\Lambda \phi_2$ as  gauge fixing condition
is
equivalent, at the level of the metric, to the  rescaling $\psi\to
(Q+\L)\psi$. Undoing this rescaling takes the resulting
 background into that of
$T^{1,Q}$ presented above.

The central charge of this model is (see Eq. (\ref{cc}))
\begin{equation}
c=  {3k_1\over k_1+2}+{3k_2\over k_2+2}-1=
   {3k\over k+2}+{3kQ^2\over kQ^2+2}-1 \ ,
\end{equation}
and reduces to 5 in  the semiclassical limit $(k\to \infty)$.
 In order to
get  a critical string background we need to    add another model
to compensate for the central charge deficit.
One  natural possibility  is to
consider a  Lorentzian version of $T^{p,q}$.
Namely,   consider
the GMM
model for  $G=SL(2,R)$, \, $G'=SL(2,R)$ and $H=U(1)$. The group
elements are
parametrized as
\begin{eqnarray}
U&=&\exp(i\phi_1\s_3)\exp(r_1\s_2)\exp(i\psi_1\s_3), \nonumber \\
V&=&\exp(i\p_2\s_3)\exp(r_2\s_2)\exp(i\psi_2\s_3)
\end{eqnarray}
and by analogy with the  $SU(2) \times SU(2)$
case
 we  define  the following
action of the subgroup
\begin{equation}
\psi_1\to \psi_1-p\epsilon(z,\bz), \qquad \p_2\to \p_2+q\epsilon(z,\bz).
\end{equation}
Following  the same steps as  above   we end up with the following background
$$
ds^2=k \bigg[dr_1^2+\sinh^2 r_1 d\p_1^2 + Q^2(dr_2^2+\sinh^2r_2 d\p_2^2)
 $$
\begin{equation}
-\ (dt+\cosh r_1 d\p_1+Q\cosh r_2 d\p_2)^2\bigg]  \ ,
\end{equation}
\begin{equation}
B_{\p_1 t}= k\cosh r_1, \quad B_{\p_1\p_2}=kQ\cosh r_1\cosh r_2, \quad
B_{\p_2 t}=-kQ\cosh r_2\ .
\end{equation}
This metric (which may be viewed as a formal
analytic continuation of the above $T^{1,Q}$ metric (3.5))  belongs to a class of
noncompact versions of Stiefel manifold,
and
corresponds to $W_{4,2}=SO(2,2)/SO(2)$.
 The parameters $k,Q$  here   are the same as above,
 so that the deficit of the central charge cancels
 just as in the $SL(2,R) \times SU(2)$ WZW model
 (to  the leading approximation in the bosonic case, and exactly
 in the supersymmetric case).

One can check that the total $d=10$  background we constructed  is not
supersymmetric. This is in contrast
to what happens in the case  of $W_{4,2}\times
T^{1,1}$  Freund-Rubin type solution of  IIB supergravity
supported by 5-form field \cite{stro}, where  the metrics on the cosets
 are chosen to be the   Einstein ones, and  1/4 of
supersymmetry is preserved.


\section{Check  of conformal invariance}

It is easy to check that the  one-loop conformal invariance equations
$R_{\mu\nu}    - { 1 \over  4} H_{\m\lambda\rho} H_{\nu}^{\ \
\lambda\rho}
=0$ \cite{callan, braaten} are satisfied. The     components of the
Ricci tensor and the scalar curvature of the $T^{1,Q}$ sector are
\begin{eqnarray}
R_{\te_1\te_1}&=&{1\over 2}, \quad R_{\p_1\p_1}={1\over
2Q^2}(Q^2+\cos^2\te_1), \quad R_{\p_1\psi}={Q^2+1\over 2Q^2} \cos
\te_1, \nonumber \\
R_{\te_2\te_2}&=&{1\over 2}, \quad R_{\p_2\p_2}={1\over
2}(1+Q^2\cos^2\te_2), \quad R_{\p_2\psi}={Q^2+1\over 2Q} \cos
\te_2, \nonumber \\
R_{\psi\psi}&=&{Q^2+1 \over 2Q^2}, \quad R_{\p_1\p_2}={Q^2+1 \over
2Q}\cos\te_1\cos\te_2, \quad R={3\over 2}{Q^2+1\over kQ^2},\nonumber \\
\end{eqnarray}
and similar expressions are found  for $ W_{4,2}$.
The total scalar curvature of
is {\it zero}
 since $R(W_{4,2})=-R(T^{1,Q}) =- 3(Q^2+1)/(2kQ^2)$.

 The
two-loop $\beta$-function for the $G_{\mu\nu} + B_{\mu\nu}$
coupling  of the bosonic
 sigma model is
(assuming  a specific scheme, see \cite{2loops, hull}):
\begin{eqnarray}
\beta_{\m\n} &=&\a'\hR_{\m\n}+{\a'^2\over
2}\left[\hR^{\a\b\g}{}_{\n}\hR_{\m\a\b\g} -{1\over
2}\hR^{\b\g\a}{}_{\n}\hR_{\m\a\b\g}+
{1\over 2}\hR_{\a \m\n \b}(H^2)^{\a\b}\right] +O(\a'^3),
\label{2loop}
\end{eqnarray}
where $\hR_{\a\b\g\d}$  is
the Riemann tensor for  the generalized connection
$\hat \G^\l{}_{\m\n}=\G^\l{}_{\m\n}-{1\over 2}H^\l{}_{\m\n}$.
In this scheme a parallelizable manifold having $\hR_{\a\b\g\d}=0$
(e.g. a group space)
 automatically
satisfies the two-loop conformal invariance condition.
 For the background we are
discussing
 the tensor  $\hR_{\a\b\g\d}$
 is  non-vanishing;  e.g.,
   the  generalized curvature of  the $T^{1,Q}$ metric is
\begin{equation}
\hR_{\te_2\p_2\te_1\p_1}=-kQ\sin\te_1\sin\te_2\  .
\end{equation}
One can check, however, that the beta-function  (\ref{2loop})
still vanishes.\footnote{Note again  that the one-loop beta function
equal to  the
generalized Ricci tensor $\hat R_{\mu\nu}$
vanishes  since $g^{\te_1\te_2}=g^{\p_2\p_1}=
g^{\te_2\p_1}= g^{\te_1\p_2}=0$.}
Like  target space backgrounds  appearing in the
case of gauged WZW models \cite{wil,2loopgwzw},
 these backgrounds, though  not parallelizable,
define conformal sigma models.

\section*{Acknowledgments}

We are grateful to K. Sfetsos for comments. LAPZ would
like to acknowledge the Office of the Provost at the University of
Michigan and the High Energy Physics Division of the Department of
Energy for support. The work of AAT  was  supported in part by
the DOE grant DOE/ER/01545,
EC TMR grant ERBFMRX-CT96-0045,
INTAS project 991590
and PPARC SPG grant  PPA/G/S/1998/00613.
He also acknowledges  the hospitality of the Schr\"odinger Institute
for Mathematical Physics  in Vienna  where this paper was completed.


\end{document}